\documentclass[twocolumn,prb,amsmath,amssymb,superscriptaddress,floatfix,showpacs]{revtex4} 

\usepackage{graphicx}
\usepackage{dcolumn}
\usepackage{bm}

\begin{document}
\setlength{\topmargin}{-0.6in}

\title{High-energy magnetic excitations from dynamic stripes in 
La$_{1.875}$Ba$_{0.125}$CuO$_4$}
\author{Guangyong Xu}
\author{J. M. Tranquada}
\affiliation{Condensed Matter Physics \&\ Materials Science Department, 
Brookhaven National Laboratory, Upton, New York 11973, USA}
\author{T. G. Perring}
\affiliation{ISIS Facility, Rutherford Appleton Laboratory, Chilton, Didcot, 
Oxon, OX11 0Qx, UK}
\author{G. D. Gu}
\affiliation{Condensed Matter Physics \&\ Materials Science Department, 
Brookhaven National Laboratory, Upton, New York 11973, USA} 
\author{M. Fujita}
\author{K. Yamada}
\affiliation{Institute for Materials Research, Tohoku University, Sendai,
980-8577, Japan}

\date{\today} 
 
\begin{abstract}

We use inelastic neutron scattering to study the temperature dependence 
of magnetic excitations (for energies up to 100 meV) in the cuprate 
La$_{1.875}$Ba$_{0.125}$CuO$_4$.  This compound exhibits stripe order
below a temperature of $\sim$50~K; previous measurements have shown
that the magnetic excitations of the stripe-ordered phase have an
hour-glass-like dispersion, with a saddle point at $\sim50$~meV.  Here
we compare measurements in the disordered phase taken at 65 and 300~K. 
At energies on the scale of $k_{\rm B}T$, there is substantial
momentum-broadening of the signal, and the low-energy incommensurate
features can no longer be resolved at 300~K.  In contrast, there is
remarkably little change in the scattered signal for energies greater
than $k_{\rm B}T$.  In fact, the momentum-integrated dynamic
susceptibility is almost constant with temperature.  We suggest that the
continuity of higher-energy magnetic excitations is strong evidence for
dynamic stripes in the high-temperature, disordered phase.  We reconsider
the nature of the magnetic dispersion, and we discuss the correspondences
between the thermal evolution of magnetic stripe correlations with other
electronic properties of the cuprates.

\end{abstract}


\maketitle

\section{Introduction}

Antiferromagnetic interactions are commonly believed to play a crucial
role in the mechanism of high-temperature
superconductivity in the cuprates.\cite{kast98,oren00} While the
long-range N\'eel order of the insulating parent compounds is suppressed
when a small density of holes is doped into the planes, dynamic spin 
correlations nevertheless persist into the superconducting phase.  A
central debate concerns the origin of the magnetic excitations.  They have
been attributed both to remnants of the parent insulator due to
segregation of the doped holes into
``stripes''\cite{kive03,zaan01,mach89,sach91} and, alternatively, to
excitations of the charge carriers across the Fermi surface, with
enhancements due to electronic
interactions.\cite{brin99,kao00,yama01,chub01,onuf02} 

Recent inelastic neutron scattering studies have made it clear that the
magnetic excitations from various cuprate families, such as 
La$_{2-x}$Sr$_x$CuO$_4$
(LSCO)\cite{maso96,lake99,tran04b,gila04,chri04,vign07} and
YBa$_2$Cu$_3$O$_{6+x}$ (YBCO),\cite{hayd04,rezn04,stoc05,pail04,hink04}
share similar traits. Many of the observed features can  be described
with a phenomenological model involving a universal magnetic  excitation
spectrum multiplied by a spin-gap function.\cite{tran06a} The onset of
the spin gap below the superconducting transition temperature, $T_c$,
together with the shift in spectral weight from below to above the gap,
can qualitatively reproduce the commensurate ``resonance'' peak  observed
in YBCO~\cite{bour98,dai01} and Bi$_2$Sr$_2$CaCu$_2$O$_{8+\delta}$ 
(BSCCO),\cite{fong99} and the incommensurate resonance in
LSCO,\cite{maso96,tran04b,chri04} when these systems go into the
superconducting state.  The magnitude of the spin gap is material
dependent and correlated with $T_c$.\cite{tran05c}  Given these universal
trends, it is reasonable to anticipate that the electronic correlations
underlying the magnetic response must be common to the various cuprate
families.

The sample studied here, La$_{1.875}$Ba$_{0.125}$CuO$_4$ [LBCO
($x=1/8$)], is a compound in which
superconductivity is anomalously suppressed ($T_c < 6$~K for our
sample\cite{tran04}) by spin and charge stripe ordering. The existence of
static spin and charge  stripes at temperatures below the structural
transition at $T_{\rm st}\sim 54$~K has been confirmed by
neutron,\cite{fuji04} resonant soft-X-ray,\cite{abba05} and
hard-X-ray\cite{huckun} diffraction measurements.  Further evidence of
the spin order (below 40~K) comes from magnetization\cite{huck05b} and
muon spin-rotation\cite{savi05} studies.  The high-energy excitation
spectrum from LBCO ($x=1/8$) in the stripe ordered state has been
established in our previous experiment.\cite{tran04}  Unlike the
isostructural nickelate compound La$_{2-x}$Sr$_x$NiO$_4$, where
spin-wave-like excitations are observed in association with diagonal
stripe order,\cite{boot03,bour03,woo05} the magnetic spectrum from our
sample is very similar to that observed in other high-$T_c$
cuprates,\cite{chri04,hayd04,rezn04,stoc05} where static stripes are
absent.   The spectrum was found to disperse only inwards from the
incommensurate  elastic peaks, and to merge at the antiferromagnetic wave
vector at an energy of $\sim50$~meV.  Above this saddle-point energy, the
excitations disperse outward from the commensurate position again, but
with an apparent 45$^\circ$ rotation of the anisotropic intensity pattern
for wave vectors in the
$a$--$b$ plane, compared to the pattern at energies below
the saddle point. 

The fact that LBCO ($x=1/8$) exhibits a stripe-ordered phase makes it an
exception compared to the other cuprates studied so far, in the sense that
there is little ambiguity on the origin of the  incommensurate magnetic
response.  Consequently, the remarkable similarities of the magnetic
excitations among these various systems suggest that stripe correlations
represent a prime candidate for explaining the universal behavior.  Of
course, if the concept of dynamic stripes is to be taken seriously, it is
important to fully characterize the magnetic excitations of a system
where a good case for dynamic stripes can be made.  A previous
study\cite{fuji04} of LBCO ($x=1/8$) focused on low-energy excitations
($\le 12$~meV) and demonstrated a continuous evolution of the magnetic
correlations across
$T_{\rm st}$, consistent with a transition from static to dynamic stripes.

Here we present an investigation of the high-energy magnetic
excitations (up to $\sim100$~meV) in LBCO ($x=1/8$) at several different
temperatures using a time-of-flight neutron spectrometer.  In
particular, we compare measurements at 65~K and 300~K, where there is no
stripe order, with those in the ordered phase at 10~K. A thermal
broadening of the momentum ({\bf Q}) dependence and  reduction of the
incommensurability is observed with increasing temperature in the
low-energy part of the spectrum  ($\alt 20$~meV), consistent with previous
work\cite{fuji04}; however, the excitations near and above the 
saddle-point energy ($\sim50$~meV) show remarkably little change.  After
describing the experimental details in Sec.~II and presenting the results
in Sec.~III, we will argue in Sec.~IV that they provide strong evidence
for, and a signature of, dynamic stripes. We will discuss the
relationship between the thermal evolution of the spin fluctuations and
other electronic properties of the cuprates.

\section{Experiment}

Our sample consists of four single 
crystals of LBCO ($x=1/8$) with total mass of 58~g; these are the same
crystals used in Ref.~\onlinecite{tran04}. The crystals were grown by the
traveling-solvent  floating-zone method at Brookhaven National
Laboratory.   At room temperature, LBCO ($x=1/8$) is in the
high-temperature-tetragonal (HTT) phase (space group $I4/mmm$); it
transforms to the low-temperature-orthorhombic (LTO) phase (space group
$Bmab$) at $\sim235$~K, and to the low-temperature-tetragonal (LTT) phase
(space group $P4_2/ncm$) at $T_{\rm st}\approx54$~K
(Ref.~\onlinecite{huck05b}). As the small changes in lattice parameters
are well below the resolution of the present measurements, we will use
the high-temperature-tetragonal unit cell, with
$a=b=3.78$~\AA\ and
$c=13.2$~\AA, to describe our results. The crystals are co-aligned
together so that the effective mosaic is
$\alt 1^\circ$ in the $a$--$c$ plane and $\alt 2^\circ$ in the  $b$--$c$
plane. 

$T_c$, measured with magnetic susceptibility on pieces cut from the ends
of our crystals, is generally less than 6~K.  Other characterizations
have recently been performed on sister crystals. 
Measurements\cite{home06} of the in-plane resistivity and optical
conductivity for $T>T_{\rm st}$ are quantitatively similar to results for
LSCO at the same hole concentration.  For $T<T_{\rm st}$, the optical
conductivity, and also angle-resolved photoemission spectra,\cite{vall06}
provide evidence for a $d$-wave-like gap.

Inelastic neutron scattering experiments have been performed on the 
MAPS time-of-flight spectrometer at the ISIS facility, Rutherford
Appleton Laboratory. For all measurements, the $c$-axis was oriented
parallel to the incident neutron  beam. Data were collected with incident
energies of $E_i = 80$~meV and $140$~meV (and chopper frequencies of
300~Hz and 400~Hz, respectively).  For the data presented here, we have
taken advantage of the four-fold crystalline symmetry to combine data
from four equivalent Brillouin zones (all at the same $|{\bf Q}|$) into
one, thus improving the statistics.  The scattered intensity has been
converted to absolute units by normalization to measurements of diffuse
scattering from a vanadium sample.

The differential scattering cross section can be written as
\begin{equation}
 \frac{d \sigma}{d\Omega_f dE_f} = \frac{k_f}{k_i}
 \tilde{S}({\bf Q},\omega),
\end{equation}
where $\tilde{S}({\bf Q}, \omega)$ is the dynamic structure factor, with
units of mbarn/(steradian meV formula unit), and
$k_i$ and $k_f$ are the initial and final neutron wave vectors. For
magnetic correlations, one has
\begin{equation}
 \tilde{S}({\bf Q},\omega) = \left(\frac{\gamma r_0}{2}\right)^2 
  f^2({\bf Q}) S({\bf Q}, \omega),
\end{equation}
where $f({\bf Q})$ is the magnetic form factor
for Cu$^{2+}$, $S({\bf Q}, \omega)$ is the Fourier transform of the
spin-spin correlation function, and $(\gamma r_0/2)^2 = 72.6$~mbarn.  The
function $S({\bf Q}, \omega)$ is related to the imaginary part of the
dynamical susceptibility by the formula
\begin{equation}
  S({\bf Q},\omega) = \chi''({\bf Q},\omega) \left/
  \left(1-e^{-\hbar\omega/k_{\rm B}T}\right)\right..
\end{equation}

The magnetic form factor for Cu$^{2+}$ is known to be
anisotropic,\cite{sham93} falling off more slowly for ${\bf Q}_\bot$,
perpendicular to the CuO$_2$ planes, than for ${\bf Q}_\|$.  For the
chosen sample orientation (${\bf c}\parallel{\bf k}_i$), we can collect
data as a function of ${\bf Q}_\|$ and $\hbar\omega$; however, the value
of ${\bf Q}_\bot$ varies with ${\bf Q}_\|$ and $\hbar\omega$, and it also
depends on $E_i$.  We have taken into account all three components of
{\bf Q} when correcting for the magnetic form factor to obtain $S$ and
$\chi''$.

\section{Results and Analysis}

\begin{figure*}[ht]
\includegraphics[angle=90,width=\linewidth]{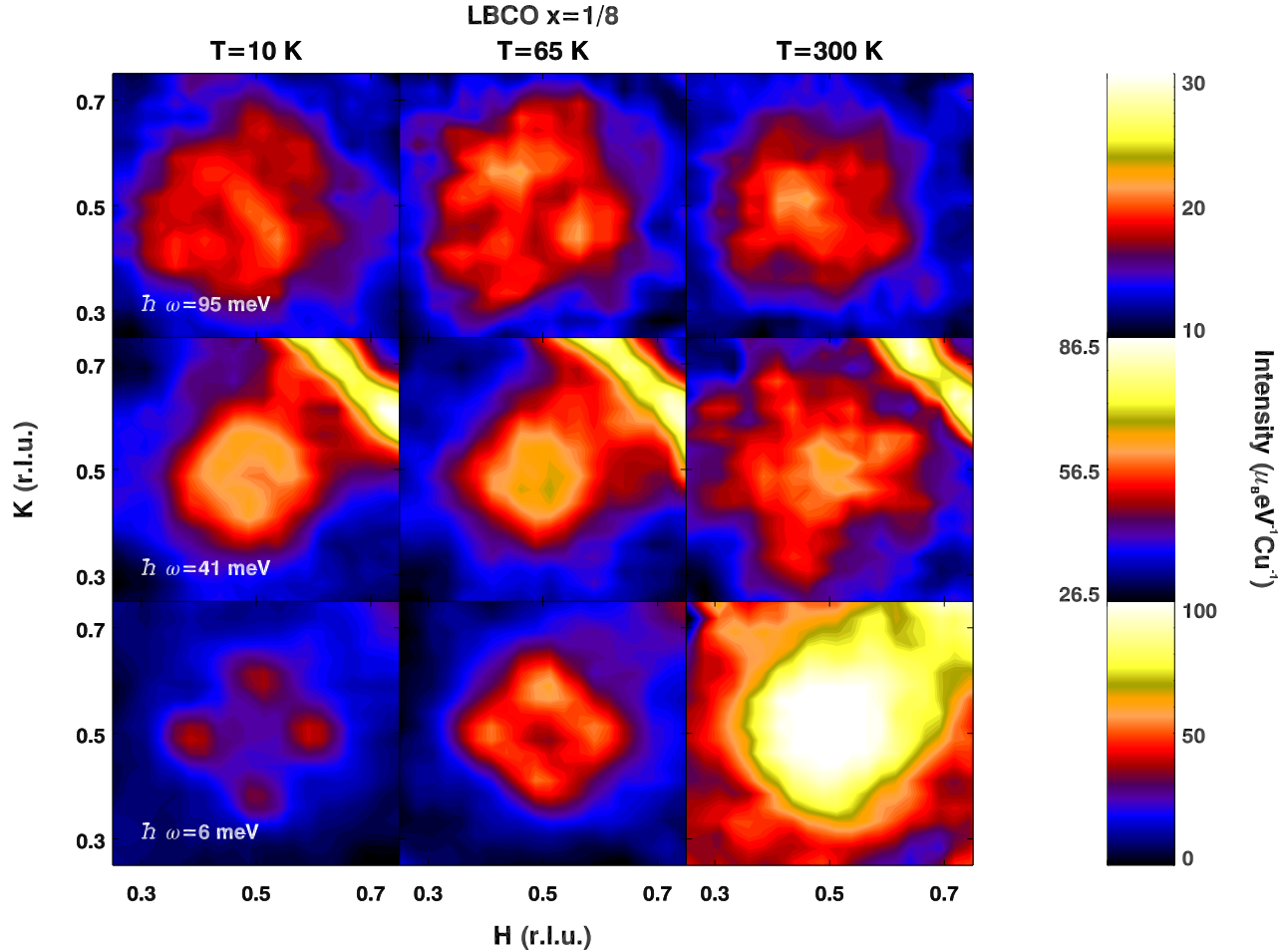}
\caption{(Color online) Constant energy cuts around ${\bf Q}_{\rm AF} =
(0.5,0.5)$ from data taken on MAPS at  different temperatures. The bottom
and middle rows are date  taken with E$_i$=80 meV, and the top row  are
data taken with E$_i$=140 meV.  } 
\label{fig:1}
\end{figure*}

Figure~\ref{fig:1} shows constant-energy slices of the scattering near the
antiferromagnetic wave vector, ${\bf Q}_{\rm AF} = (0.5,0.5)$, for three
temperatures.  The energy increases (6, 41, and 95 meV) from bottom to
top, and the temperature increases (10, 65, and 300~K) from left to
right.  The intensity peaked near ${\bf Q}_{\rm AF}$ we attribute to
magnetic scattering; the ``background'' signal is due mostly to phonons. 
The bright diagonal stripe of scattering in the upper-right corner of the
panels for 41~meV is also attributed to phonons.

The incommensurate peaks characteristic of stripe
correlations\cite{tran95a,fuji04} are clearly seen in the
$\hbar\omega=6$~meV data at 10 and 65~K; separate peaks are no longer
resolved at 300 K.  Note that static stripe order is present at 10~K, but
is absent at 65 and 300 K, where $T>T_{\rm st}$.  As observed
previously,\cite{fuji04} the loss of stripe order results in a small
decrease in the peak incommensurability and significant
momentum-broadening of the scattered intensity.  

In striking contrast, there is very little change in the high-energy
scattering (41 and 95 meV) across $T_{\rm st}$.  By looking at the
high-energy data, it is impossible to judge whether or not stripe-order
is present.  Admittedly, at 95 meV the shape of the magnetic
scattering is limited by counting statistics; nevertheless, it is clear
that there are no dramatic changes in the intensity or general {\bf Q}
dependence of the magnetic signal.  Even at 300~K, a temperature roughly
six times the stripe-ordering temperature, the main effect at 41~meV seems
to be relatively modest momentum broadening of the magnetic signal.  

\begin{figure}[ht]
\includegraphics[angle=90,width=\linewidth]{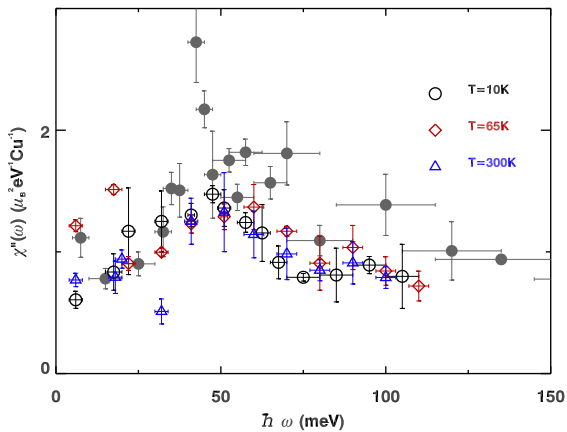}
\caption{(Color online) Imaginary part of the spin susceptibility, after
{\bf Q} integration over the antiferromagnetic Brillouin zone,
$\chi''(\omega)$ measured at T=10~K (black open circles), 65~K (red
diamonds), and 300~K (blue triangles). The solid gray circles
represent data from Ref.~\onlinecite{tran04}.  } 
\label{fig:2}
\end{figure}

Another way to examine the thermal evolution of the magnetic correlations
is through the density of states.  Figure~\ref{fig:2} shows
$\chi''(\omega)$, the imaginary part of the spin susceptibility
integrated over {\bf Q} in the antiferromagnetic Brillouin zone.  The
biggest contribution to the error bars comes from uncertainties in
subtracting background contributions.  The attentive reader will note
that the sharp spike at $\sim45$~meV that was present in the previous
results~\cite{tran04} is absent in the new results.  We are now confident
that this feature is due to a dispersive phonon, as it is also
present~\cite{woo05} in isostructural La$_{1.67}$Sr$_{0.33}$NiO$_4$; we
have taken better account of this phonon contribution in the current
analysis. Background contributions including those 
from higher energy phonons ($\agt 50$~meV) are also more carefully examined, 
and the resulting $\chi''(\omega)$ is slightly smaller than, but within error
bars of that from previous analysis.

We have chosen to present $\chi''(\omega)$, rather than $S(\omega)$,
because it more clearly illustrates the lack of dramatic changes with
temperature.  The lack of significant change is especially clear for
$\hbar\omega > 30$~meV.  Below 20 meV, we find an initial increase in
$\chi''(\omega)$ on losing static stripe order, and then a return to
lower values at 300~K.  This behavior is somewhat different from that
inferred from measurements of low-energy spin excitations with a
triple-axis spectrometer.\cite{fuji04}  There, fitting of a particular
model cross section to the one-dimensional scans in {\bf Q} indicated
that $\chi''(\omega)$ monotonically decreased with warming from the
ordered state.  The differences could possibly result from inadequacy of
the model cross section or differing treatments of background, but we
cannot rule out actual differences in the data.

The main point here is that over a substantial energy range there is
little temperature dependence to $\chi''(\omega)$.  Such behavior is what
one typically finds in an ordered system, where the only impact of
temperature is to change the occupancy of excited states.  It is
reminiscent of the variation of the scattering from the
antiferromagnetically-correlated spins in La$_2$CuO$_4$ on warming
through the N\'eel temperature.\cite{yama89}  There, again, the loss of
order is not readily detectable at higher excitations energies.

\begin{figure}[ht]
\includegraphics[width=\linewidth]{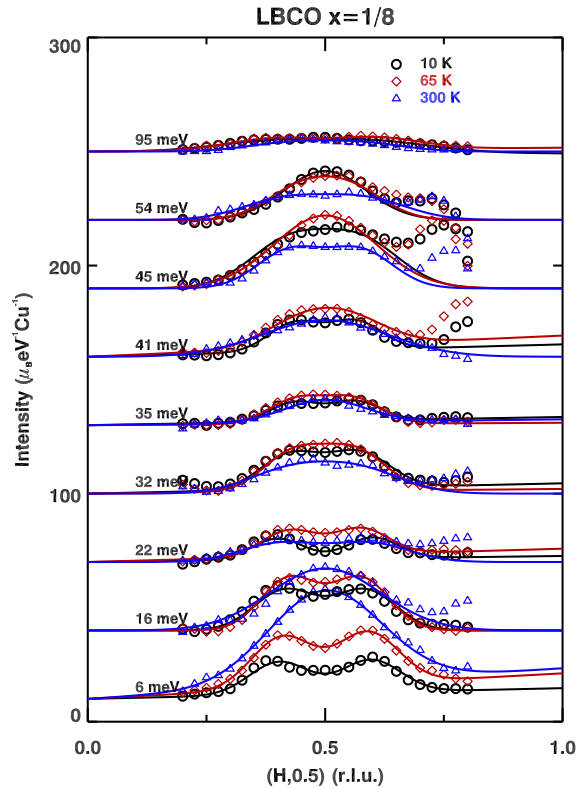}
\caption{(Color online) Linear intensity profiles measured through 
${\bf Q}_{\rm AF}$ at different energy  transfers. For each data set, we
average linear cuts taken along [100] and  [010] directions to improve
statistics. The solid lines are fits to  the data using two symmetric
Gaussians displaced equally from $H=0.5$, plus background. The peaks on
the high-$Q$ side (e.g., at $\hbar\omega=16$~meV, 32~meV, 41~meV, and
45~meV) are due to phonon contributions.} 
\label{fig:3}
\end{figure}

Returning to the {\bf Q} dependence of the data, one-dimensional
constant-$E$ cuts through ${\bf Q}_{\rm AF}$, taken along [100] and [010]
directions and averaged, are plotted in Fig.~\ref{fig:3}.  (In the {\bf Q}
direction perpendicular to each cut, data were averaged over a gaussion
window with gaussian half-width, $\sigma_q = 0.02$~r.l.u.)  We analyze
each scan in terms of a symmetric pair of gaussian peaks split about
${\bf Q}_{\rm AF}$ and sitting on top of a weakly
temperature-dependent background.  At low temperature, the results are
consistent with our previous observations.\cite{fuji04,tran04}  We can
cleanly resolve incommensurate peaks in scans up to about 30 meV.  At
higher energies, the peaks move closer together, forming one broad peak
that reaches a minimum width at $\sim54$~meV.  The scattering broadens and
weakens at still higher energies. 

With increasing temperature, the peak incommensurability (at low
energies) is slightly reduced and the peak widths increase, consistent
with previous work.\cite{fuji04}  The increase in $Q$ width indicates a
decrease in the spin-spin correlation length.\cite{fuji04,aepp97}  At
65~K, the incommensurability is less well resolved than in the ordered
state, and no incommensurability can be detected for $T=300$~K.  

\begin{figure}[ht]
\includegraphics[angle=90,width=\linewidth]{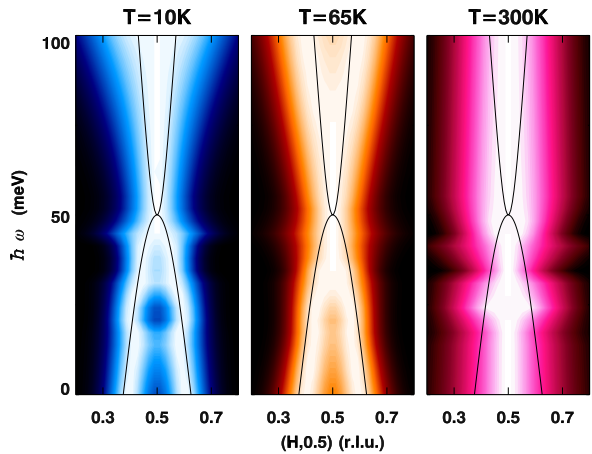}
\caption{(Color online) Two-dimensional maps plotted 
in energy-momentum space showing the fitted dispersions and $Q$ widths at
$T=10$~K, 65~K, and 300~K. The signal is sliced along the [100]
direction in ${\bf Q}$, through ${\bf Q}_{\rm AF}$. As a reference, the
solid lines denote the dispersion at low temperature obtained in
Ref.~\onlinecite{tran04}. } 
\label{fig:4}
\end{figure}

To summarize the energy and temperature dependence of the peak positions
and widths, we have used the parameters obtained in the gaussian fits
(indicated in Fig.~\ref{fig:3}) to create the contour plots shown in
Fig.~\ref{fig:4}; here the peak intensity at each energy has been
normalized to one.  The solid lines indicate the effective dispersion
characterizing our previous results\cite{tran04} at 10~K. 
Qualitatively, the dispersion at $T=10$~K (stripe-ordered phase) and at
$T=65$~K (disordered phase) are the same.  At $T=300$~K, while there is
no incommensurability, the overall $Q$ width is not greater than the
spread in $Q$ (due to incommensurability at low energies and upward
dispersion at higher energies) at the lower temperatures. 

The instantaneous spin-spin correlations are described by the equal-time
spin correlation function, 
$$ S({\bf Q})=\int S({\bf Q},\omega)d\omega.$$ 
It is fairly clear from the data in Fig.~\ref{fig:3} that, in the
disordered phase, $S({\bf Q})$ will be a broad peak centered at
${\bf Q}_{\rm AF}$.  We have used two methods to evaluate the integral. 
The first approach is to directly integrate the experimental $S({\bf
Q},\omega)$ data.  There are two drawbacks to this method.  One is that
the dominant signal is due to phonons (see Fig.~1 of
Ref.~\onlinecite{waki06}); integration over energy should give some
averaging over the phonon signals, but is not guaranteed to yield a smooth
background.  The second issue is the magnetic form factor, which varies
with energy because $Q_z$, the component of {\bf Q} perpendicular to the
CuO$_2$ planes, is a function of the energy transfer.  In dividing the
form factor out of the data, we also modify the phonon contributions
whose {\bf Q} dependence we are hoping will largely cancel in the
integral.  This correction did not seem to have much impact on the nature
of the net phonon background.  The second method used is to integrate
signal obtained from fits to the data, as shown in Fig.~\ref{fig:3}.

\begin{figure}[t]
\includegraphics[width=\linewidth]{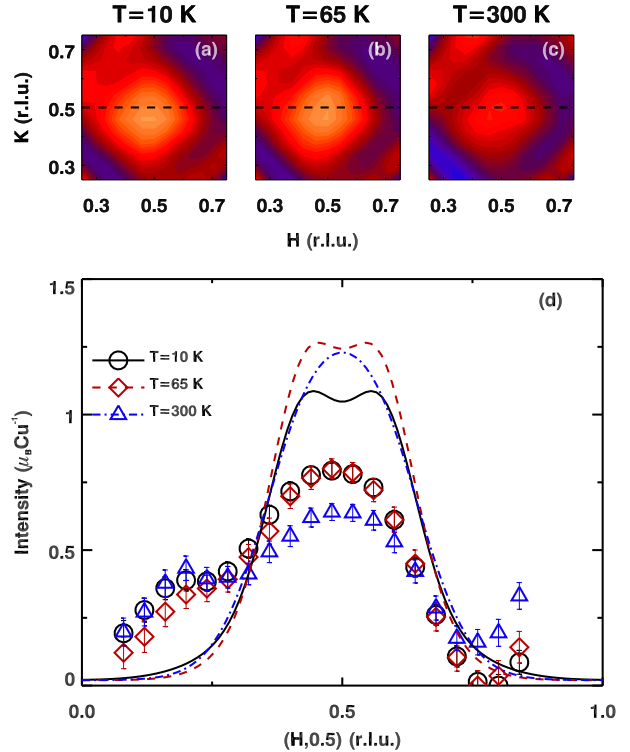}
\caption{(Color online) 
Measurements of $S({\bf Q})$.  (a), (b), and (c) show the energy
integrals (over the interval 4 to 110 meV) of the $E_i=140$~meV data,
corrected for the magnetic form factor, at temperatures of 10~K, 65~K,
and 300~K, respectively.  A background varying as $Q^2$ has been
subtracted in order to emphasize the magnetic response.  The diagonal
streaks near the upper left and lower right corners are artifacts due to
spurious signals at the detector edges.  (d) Cuts along the {\bf Q} path
indicated by dashed lines in (a)-(c).  Symbols represent the integral of
over the data, while the lines correspond to integration over the fitted
magnetic signal (from Fig.~\ref{fig:3}).  The difference in the two
approaches is due to the differing treatments of the phonon background.} 
\label{fig:5}
\end{figure}

The results for $S({\bf Q})$ are presented in Fig.~\ref{fig:5}.  Panels
(a)-(c) show the results obtained from direct integration of the data,
after subtraction of a background proportional to $Q^2$, at temperatures
at 10~K, 65~K, and 300~K.  One can see that the shape and size of the
instantaneous structure factor does not change greatly with temperature. 
Cuts along ${\bf Q} = (H,0,0)$ are shown in Fig.~\ref{fig:5}(d).  The
symbols represent the direct integration over the data, while the lines
correspond to the integration over the fitted magnetic signal.  The
difference in the results from the two approaches is due to the limited
cancellation of phonon structures in the direct integration method.  We
believe that the integration over the fits provides a more reliable
measure of $S({\bf Q})$.  We should also note that the result at 10~K is
incomplete in that we have not included the sharp peaks from the elastic
channel in the integration; although the relative weight in the elastic
peaks is not large, their narrow $Q$ widths would lead them to dominate
the shape of $S({\bf Q})$.

From Fig.~\ref{fig:5}(d), one can see that the width of $S({\bf Q})$
shows virtually no change with temperature. Quantitatively, the half width
is $\kappa = 0.15\pm0.02$~r.l.u.  If we were to assume that the spin
correlations fall off exponentially in real space, then we would obtain
a correlation length of $\xi = 2\pi/\kappa a = 4.0\pm0.5$~\AA, or
approximately one Cu-Cu nearest-neighbor spacing.  The values we obtain
for $\kappa$ and $\xi$ are virtually the same as those reported by Hayden
{\it et al.}\cite{hayd96a} for LSCO $x=0.14$ at 17~K; the magnitude of
$S({\bf Q})$ is also similar.  

In considering the significance of $\kappa$ and its lack of temperature
dependence, it is worth noting that the shape of $S({\bf Q})$ (at least
that given by the lines in Fig.~\ref{fig:5}) is different from the
lorentzian one would expect if the spatial correlations decayed
exponentially.  Instead, the shape is rather flat-topped.  To see what
may determine this shape, it is instructive to look back at
Fig.~\ref{fig:4}.  The limiting feature for the $Q$ width seems to be the
low-energy incommensurability that is resolvable at lower temperatures. 
That feature is determined by the facts that the antiferromagnetic
correlations are constrained to narrow domains by the charge stripes
and that the phase of the antiferromagnetic correlations must flip by
$\pi$ on crossing a charge stripe.  Thus, the magnitude of $\kappa$,
together with the large energy scale for the magnetic excitations, appear
to reflect the spatial inhomogeneity of the spin and charge correlations.

\section{Discussion}

The present results show that the energy scale characterizing spin
fluctuations in LBCO ($x=\frac18$) is insensitive to the ordering of
charge stripes.  Given that the maximum in $\chi''(\omega)$ is at
$\sim50$~meV, it is not surprising that thermal energy alone has a modest
impact on the spectrum, even at 300~K.  What is more important is the
observation that there is no qualitative change in the magnetic spectrum
due to the loss of stripe order.  This is another piece of evidence that
the electronic and magnetic correlations in the stripe-ordered phase are
similar to those in the disordered phase.  It is consistent with the idea
that the disordered state consists of a stripe liquid\cite{zaan96a} or
nematic phase.\cite{kive98}  Evidence for a stripe liquid phase has also
been observed in layered nickelates\cite{lee02}; the difference in
cuprates is that quantum fluctuations may prevent ordering, in the
absence of a crystal symmetry that removes the degeneracy between
equivalent stripe orientations.

That the electronic properties of the stripe-ordered state are subtly
different from those in the disordered phase has been demonstrated by a
couple of recent experiments. Angle-resolved photoemission and tunneling
measurements\cite{vall06} indicate a d-wave-like gap similar to that
found in superconducting samples.  The maximum amplitude of the gap is
$\sim20$~meV.  The gap also appears in the in-plane optical
conductivity,\cite{home06} where one sees evidence both for excitations
across the gap and a collective mode.  In the disordered phase, the
in-plane optical conductivity looks essentially identical to that in LSCO
and other superconducting cuprates of comparable hole content.

A picture of alternating doped and undoped two-leg ladders, with no
ordered spins in the doped ladders, is consistent with the magnetic stripe
order.\cite{fuji04}  It was argued previously\cite{tran04} that the
effective magnetic dispersion observed in the stripe-ordered phase can be
understood in terms of the excitations of the ordered spin ladders with a
weak effective coupling across the doped ladders; the energy
scale is set by the superexchange energy of the parent insulator
(La$_2$CuO$_4$ in this case).\cite{tran04,tran06a,tran05c}  Various model
calculations of the excitation spectra have been based on this
picture.\cite{vojt04,uhri04,vojt06,yao06,ande06}  It should be noted,
however, that this is certainly an over-simplified starting point for the
spin dynamics, because it ignores the spin degrees of freedom in the doped
stripes.  Here we reconsider the nature of the dispersion.

Inelastic neutron scattering measurements on under-doped crystals of YBCO
with varying degrees of twinning indicate that the magnetic excitations
above the ``resonance'' energy seem to have an isotropic dispersion and
intensity, in contrast to excitations at lower energies, which do show
significant orientational anisotropy.\cite{stoc05,hink06}  These YBCO
crystals do not exhibit any static stripe order; however, the dispersions
do show strong similarities to those in LBCO,\cite{tran05c} and it seems
sensible to look for a common understanding.  From the similar
characters, we infer that the magnetic excitations above 50~meV in LBCO
likely close to 4-fold symmetric within each CuO$_2$ plane.  
Such behavior would be inconsistent with the over-simplified 
model,\cite{vojt04,uhri04,vojt06,yao06,ande06} but could it be compatible
with a more general stripe model?

To answer this question, we first need to consider the nature of spin
excitations in doped two-leg ladders.  One relevant model system is
Sr$_{14}$Cu$_{24}$O$_{41}$, a compound containing both Cu-O chains and
lightly hole-doped\cite{pisk05,vule05,rusy06} two-leg spin ladders. 
Neutron-scattering studies\cite{boul04} of the response from the ladders
provide evidence for gapped triplet excitations very similar to those in
undoped ladders.\cite{notb07}  The
survival of singlet correlations and triplet excitations in doped two-leg
ladders is certainly expected theoretically,\cite{roux05,essl06} and the
spin correlations in such systems are believed to be relevant to hole
pairing.\cite{dago92,kive05a}

Another result relevant to the problem of doping holes into a quantum
spin system involves the Haldane-chain compound Y$_{2-x}$Ca$_x$BaNiO$_5$. 
For $x=0$, the Ni-O chains behave as decoupled $S=1$ spin chains, with a
gap of approximately 9 meV for triplet excitations\cite{xu96} and a
dispersion extending to greater than 60 meV.\cite{xu00b}  On doping holes
into the chains, through Ca substitution for Y, there is no qualitative
change in the dispersion of the triplet excitations; the biggest change
is the appearance of new incommensurate excitations within the spin gap
of the undoped system.\cite{xu00b}

With these results in mind, let us return to the case of the
stripe-ordered phase. The incommensurate magnetic peaks tell us that the
nearest-neighbor antiferromagnetic domains must have an antiphase
relationship.\cite{tran95a}  The same correlation is relevant for all of
the magnetic excitations at energies $\lesssim50$~meV.  This constraint
no longer applies at higher energies.  For
$\hbar\omega\gtrsim50$~meV, the correlated fluctuations of the moments in
neighboring domains are essentially in phase, rather than out of
phase---we have a dispersion of excitations from ${\bf Q}_{\rm AF}$ as
in an antiferromagnet with a gap.  If the gap corresponds to the the spin
gap in the doped ladders, then we might expect the spin excitations of
the doped ladders to contribute to the in-phase dispersion. The net
dispersion could be fairly isotropic.  (Note that a more isotropic
dispersion of high-energy excitations is obtained in the
Gutzwiller-mean-field calculation of stripes by Seibold and
Lorenzana,\cite{seib05} where the assumptions of the spin-only models
were not made.)  

A plausible, phenomenological picture, then, is that the magnetic
excitations above a spin gap energy [$\sim50$~meV in the case of LBCO
($x=1/8$)] are similar to two-dimensional antiferromagnetic spin
waves,\cite{stoc05} but with a very short correlation length.   The impact
of the doped holes becomes directly apparent at lower energies, where one
observes a downward dispersion.  From the evolution of the spin
excitations in LBCO from the spin-ordered to the disordered phase, the
lowest-energy incommensurate spin excitations must be the Goldstone modes
of the stripe phase.

Returning to the possibility that a stripe-liquid character is common to
the hole-doped cuprates, we note that objections have been raised in the
case of YBCO.  In particular, Bourges {\it et al.}\cite{bour00} have
pointed out that while a dispersion downward from the ``resonance'' energy
is observed in YBa$_2$Cu$_3$O$_{6.85}$ for $T<T_c$, it disappears for
$T\gtrsim T_c$.  In considering this objection, it should first be noted
that no measurements were presented for energies lower than 25 meV;
presumably, the magnetic signal was too weak to analyze at lower
energies.  Next, consider our results for LBCO at 65~K in
Fig.~\ref{fig:3}.  No incommensurate features can be resolved at energies
above $\sim30$~meV; the incommensurability only becomes clear at
lower energies.  Thus, the absence of an obvious
dispersive or incommensurate character at higher energies and
temperatures is completely compatible with our observations in the
stripe-liquid phase of LBCO.  Another relevant observation in
Figs.~\ref{fig:3} and \ref{fig:5} is that the net $Q$ width of the
dominant magnetic signal associated with the stripe correlations does not
change dramatically with temperature.  Similarly, one observes the same
net $Q$ width in YBa$_2$Cu$_3$O$_{6.85}$ both above and below
$T_c$.\cite{bour00}

In LBCO ($x=1/8$), the most significant changes with temperature seem to
be associated with the $Q$ width, $\kappa$, and incommensurability of the
low-energy excitations, as observed in a previous study.\cite{fuji04}  In
the LTO phase, $\kappa$ decreases with cooling.  This behavior is
correlated with a decrease in the energy width of the Drude peak found in
the optical conductivity.\cite{home06}  Similarly, the absence of
incommensurability in the low-energy magnetic excitations at room
temperature is correlated with the disappearance of the Drude peak into a
very broad mid-infrared continuum.  We infer that the development of
electronic coherence in the copper-oxide planes is associated with the
growth of stripe correlations on cooling.  

Such trends extend well beyond LBCO.  In optimally-doped LSCO, Aeppli
{\it et al.}\cite{aepp97} found that $\kappa$ for low-energy
incommensurate spin excitations shows quantum critical behavior,
reflecting the evolution of stripe correlations with cooling, but with 
stripe order avoided due to quantum fluctuations.  Optical conductivity
studies of LSCO show a linear variation of the Drude energy width with
temperature for a range of dopings.\cite{gao93,take02}  Thus, it appears
that, rather than competing, stripe and ``nodal-metal''\cite{lee05}
coherence develop synergistically.  It is only static stripe order that
competes with superconductivity, and even then, stripe order is
compatible with gapless electronic excitations at the nodal
points.\cite{vall06,home06}

\bigskip
\section{Conclusion}

We have presented an inelastic neutron scattering study of magnetic
excitations at energies up to 100 meV for LBCO ($x=1/8$) at temperatures
below and above the stripe ordering temperature.  While the excitations
at low energies are broadenend in $Q$ on warming, the magnetic signal for
$\hbar\omega > k_{\rm B} T$ shows remarkably little change.  We interpret
these results as evidence of a stripe-liquid phase for $T>T_{\rm
st}$.  We have noted similarities between the magnetic excitation
spectrum of the stripe-liquid phase and that seen in other cuprates.  We
have also pointed out that the coherence of the stripe-liquid
correlations and low-energy charge excitations develop synergistically
with cooling.

\section*{Acknowledgements}

It is a pleasure to acknowledge helpful discussions
with S. A. Kivelson and E. W. Carlson.  Work at Brookhaven is supported by
the U.S. Department of Energy's Office of Science under Contract
No.\ DE-AC02-98CH10886. Work at Tohoku University is supported by
a Grant-in-Aid for Scientific Research from MEXT.  
This work has also benefited from the U.S.-Japan
Cooperative Program on Neutron Scattering.


\begin{thebibliography}{69}
\expandafter\ifx\csname natexlab\endcsname\relax\def\natexlab#1{#1}\fi
\expandafter\ifx\csname bibnamefont\endcsname\relax
  \def\bibnamefont#1{#1}\fi
\expandafter\ifx\csname bibfnamefont\endcsname\relax
  \def\bibfnamefont#1{#1}\fi
\expandafter\ifx\csname citenamefont\endcsname\relax
  \def\citenamefont#1{#1}\fi
\expandafter\ifx\csname url\endcsname\relax
  \def\url#1{\texttt{#1}}\fi
\expandafter\ifx\csname urlprefix\endcsname\relax\def\urlprefix{URL }\fi
\providecommand{\bibinfo}[2]{#2}
\providecommand{\eprint}[2][]{\url{#2}}

\bibitem[{\citenamefont{Kastner et~al.}(1998)\citenamefont{Kastner, Birgeneau,
  Shirane, and Endoh}}]{kast98}
\bibinfo{author}{\bibfnamefont{M.~A.} \bibnamefont{Kastner}},
  \bibinfo{author}{\bibfnamefont{R.~J.} \bibnamefont{Birgeneau}},
  \bibinfo{author}{\bibfnamefont{G.}~\bibnamefont{Shirane}}, \bibnamefont{and}
  \bibinfo{author}{\bibfnamefont{Y.}~\bibnamefont{Endoh}},
  \bibinfo{journal}{Rev. Mod. Phys.} \textbf{\bibinfo{volume}{70}},
  \bibinfo{pages}{897} (\bibinfo{year}{1998}).

\bibitem[{\citenamefont{Orenstein and Millis}(2000)}]{oren00}
\bibinfo{author}{\bibfnamefont{J.}~\bibnamefont{Orenstein}} \bibnamefont{and}
  \bibinfo{author}{\bibfnamefont{A.~J.} \bibnamefont{Millis}},
  \bibinfo{journal}{Science} \textbf{\bibinfo{volume}{288}},
  \bibinfo{pages}{468} (\bibinfo{year}{2000}).

\bibitem[{\citenamefont{Kivelson et~al.}(2003)\citenamefont{Kivelson, Bindloss,
  Fradkin, Oganesyan, Tranquada, Kapitulnik, and Howald}}]{kive03}
\bibinfo{author}{\bibfnamefont{S.~A.} \bibnamefont{Kivelson}},
  \bibinfo{author}{\bibfnamefont{I.~P.} \bibnamefont{Bindloss}},
  \bibinfo{author}{\bibfnamefont{E.}~\bibnamefont{Fradkin}},
  \bibinfo{author}{\bibfnamefont{V.}~\bibnamefont{Oganesyan}},
  \bibinfo{author}{\bibfnamefont{J.~M.} \bibnamefont{Tranquada}},
  \bibinfo{author}{\bibfnamefont{A.}~\bibnamefont{Kapitulnik}},
  \bibnamefont{and} \bibinfo{author}{\bibfnamefont{C.}~\bibnamefont{Howald}},
  \bibinfo{journal}{Rev. Mod. Phys.} \textbf{\bibinfo{volume}{75}},
  \bibinfo{pages}{1201} (\bibinfo{year}{2003}).

\bibitem[{\citenamefont{Zaanen et~al.}(2001)\citenamefont{Zaanen, Osman, Kruis,
  Nussinov, and {Tworzyd\l o}}}]{zaan01}
\bibinfo{author}{\bibfnamefont{J.}~\bibnamefont{Zaanen}},
  \bibinfo{author}{\bibfnamefont{O.~Y.} \bibnamefont{Osman}},
  \bibinfo{author}{\bibfnamefont{H.~V.} \bibnamefont{Kruis}},
  \bibinfo{author}{\bibfnamefont{Z.}~\bibnamefont{Nussinov}}, \bibnamefont{and}
  \bibinfo{author}{\bibfnamefont{J.}~\bibnamefont{{Tworzyd\l o}}},
  \bibinfo{journal}{Phil. Mag. B} \textbf{\bibinfo{volume}{81}},
  \bibinfo{pages}{1485} (\bibinfo{year}{2001}).

\bibitem[{\citenamefont{Machida}(1989)}]{mach89}
\bibinfo{author}{\bibfnamefont{K.}~\bibnamefont{Machida}},
  \bibinfo{journal}{Physica C} \textbf{\bibinfo{volume}{158}},
  \bibinfo{pages}{192} (\bibinfo{year}{1989}).

\bibitem[{\citenamefont{Sachdev and Read}(1991)}]{sach91}
\bibinfo{author}{\bibfnamefont{S.}~\bibnamefont{Sachdev}} \bibnamefont{and}
  \bibinfo{author}{\bibfnamefont{N.}~\bibnamefont{Read}},
  \bibinfo{journal}{Int. J. Mod. Phys. B} \textbf{\bibinfo{volume}{5}},
  \bibinfo{pages}{219} (\bibinfo{year}{1991}).

\bibitem[{\citenamefont{Brinckmann and Lee}(1999)}]{brin99}
\bibinfo{author}{\bibfnamefont{J.}~\bibnamefont{Brinckmann}} \bibnamefont{and}
  \bibinfo{author}{\bibfnamefont{P.~A.} \bibnamefont{Lee}},
  \bibinfo{journal}{Phys. Rev. Lett.} \textbf{\bibinfo{volume}{82}},
  \bibinfo{pages}{2915} (\bibinfo{year}{1999}).

\bibitem[{\citenamefont{Kao et~al.}(2000)\citenamefont{Kao, Si, and
  Levin}}]{kao00}
\bibinfo{author}{\bibfnamefont{Y.-J.} \bibnamefont{Kao}},
  \bibinfo{author}{\bibfnamefont{Q.}~\bibnamefont{Si}}, \bibnamefont{and}
  \bibinfo{author}{\bibfnamefont{K.}~\bibnamefont{Levin}},
  \bibinfo{journal}{Phys. Rev. B} \textbf{\bibinfo{volume}{61}},
  \bibinfo{pages}{R11898} (\bibinfo{year}{2000}).

\bibitem[{\citenamefont{Yamase and Kohno}(2001)}]{yama01}
\bibinfo{author}{\bibfnamefont{Y.}~\bibnamefont{Yamase}} \bibnamefont{and}
  \bibinfo{author}{\bibfnamefont{H.}~\bibnamefont{Kohno}}, \bibinfo{journal}{J.
  Phys. Soc. Japan} \textbf{\bibinfo{volume}{70}}, \bibinfo{pages}{2733}
  (\bibinfo{year}{2001}).

\bibitem[{\citenamefont{Chubukov et~al.}(2001)\citenamefont{Chubukov, Jank\'o,
  and Tchernyshyov}}]{chub01}
\bibinfo{author}{\bibfnamefont{A.~V.} \bibnamefont{Chubukov}},
  \bibinfo{author}{\bibfnamefont{B.}~\bibnamefont{Jank\'o}}, \bibnamefont{and}
  \bibinfo{author}{\bibfnamefont{O.}~\bibnamefont{Tchernyshyov}},
  \bibinfo{journal}{Phys. Rev. B} \textbf{\bibinfo{volume}{63}},
  \bibinfo{pages}{180507R} (\bibinfo{year}{2001}).

\bibitem[{\citenamefont{Onufrieva and Pfeuty}(2002)}]{onuf02}
\bibinfo{author}{\bibfnamefont{F.}~\bibnamefont{Onufrieva}} \bibnamefont{and}
  \bibinfo{author}{\bibfnamefont{P.}~\bibnamefont{Pfeuty}},
  \bibinfo{journal}{Phys. Rev. B} \textbf{\bibinfo{volume}{65}},
  \bibinfo{pages}{054515} (\bibinfo{year}{2002}).

\bibitem[{\citenamefont{Mason et~al.}(1996)\citenamefont{Mason, Schr\"oder,
  Aeppli, Mook, and Hayden}}]{maso96}
\bibinfo{author}{\bibfnamefont{T.~E.} \bibnamefont{Mason}},
  \bibinfo{author}{\bibfnamefont{A.}~\bibnamefont{Schr\"oder}},
  \bibinfo{author}{\bibfnamefont{G.}~\bibnamefont{Aeppli}},
  \bibinfo{author}{\bibfnamefont{H.~A.} \bibnamefont{Mook}}, \bibnamefont{and}
  \bibinfo{author}{\bibfnamefont{S.~M.} \bibnamefont{Hayden}},
  \bibinfo{journal}{Phys. Rev. Lett.} \textbf{\bibinfo{volume}{77}},
  \bibinfo{pages}{1604} (\bibinfo{year}{1996}).

\bibitem[{\citenamefont{Lake et~al.}(1999)\citenamefont{Lake, Aeppli, Mason,
  Schr\"oder, McMorrow, Lefmann, Isshiki, Nohara, Takagi, and Hayden}}]{lake99}
\bibinfo{author}{\bibfnamefont{B.}~\bibnamefont{Lake}},
  \bibinfo{author}{\bibfnamefont{G.}~\bibnamefont{Aeppli}},
  \bibinfo{author}{\bibfnamefont{T.~E.} \bibnamefont{Mason}},
  \bibinfo{author}{\bibfnamefont{A.}~\bibnamefont{Schr\"oder}},
  \bibinfo{author}{\bibfnamefont{D.~F.} \bibnamefont{McMorrow}},
  \bibinfo{author}{\bibfnamefont{K.}~\bibnamefont{Lefmann}},
  \bibinfo{author}{\bibfnamefont{M.}~\bibnamefont{Isshiki}},
  \bibinfo{author}{\bibfnamefont{M.}~\bibnamefont{Nohara}},
  \bibinfo{author}{\bibfnamefont{H.}~\bibnamefont{Takagi}}, \bibnamefont{and}
  \bibinfo{author}{\bibfnamefont{S.~M.} \bibnamefont{Hayden}},
  \bibinfo{journal}{Nature} \textbf{\bibinfo{volume}{400}}, \bibinfo{pages}{43}
  (\bibinfo{year}{1999}).

\bibitem[{\citenamefont{Tranquada
  et~al.}(2004{\natexlab{a}})\citenamefont{Tranquada, Lee, Yamada, Lee,
  Regnault, and R{\o}nnow}}]{tran04b}
\bibinfo{author}{\bibfnamefont{J.~M.} \bibnamefont{Tranquada}},
  \bibinfo{author}{\bibfnamefont{C.~H.} \bibnamefont{Lee}},
  \bibinfo{author}{\bibfnamefont{K.}~\bibnamefont{Yamada}},
  \bibinfo{author}{\bibfnamefont{Y.~S.} \bibnamefont{Lee}},
  \bibinfo{author}{\bibfnamefont{L.~P.} \bibnamefont{Regnault}},
  \bibnamefont{and} \bibinfo{author}{\bibfnamefont{H.~M.}
  \bibnamefont{R{\o}nnow}}, \bibinfo{journal}{Phys. Rev. B}
  \textbf{\bibinfo{volume}{69}}, \bibinfo{pages}{174507}
  (\bibinfo{year}{2004}{\natexlab{a}}).

\bibitem[{\citenamefont{Gilardi et~al.}(2004)\citenamefont{Gilardi, Hiess,
  Momono, Oda, Ido, and Mesot}}]{gila04}
\bibinfo{author}{\bibfnamefont{R.}~\bibnamefont{Gilardi}},
  \bibinfo{author}{\bibfnamefont{A.}~\bibnamefont{Hiess}},
  \bibinfo{author}{\bibfnamefont{N.}~\bibnamefont{Momono}},
  \bibinfo{author}{\bibfnamefont{M.}~\bibnamefont{Oda}},
  \bibinfo{author}{\bibfnamefont{M.}~\bibnamefont{Ido}}, \bibnamefont{and}
  \bibinfo{author}{\bibfnamefont{J.}~\bibnamefont{Mesot}},
  \bibinfo{journal}{Europhys. Lett.} \textbf{\bibinfo{volume}{66}},
  \bibinfo{pages}{840} (\bibinfo{year}{2004}).

\bibitem[{\citenamefont{Christensen et~al.}(2004)\citenamefont{Christensen,
  McMorrow, R{\o}nnow, Lake, Hayden, Aeppli, Perring, Mangkorntong, Nohara, and
  Tagaki}}]{chri04}
\bibinfo{author}{\bibfnamefont{N.~B.} \bibnamefont{Christensen}},
  \bibinfo{author}{\bibfnamefont{D.~F.} \bibnamefont{McMorrow}},
  \bibinfo{author}{\bibfnamefont{H.~M.} \bibnamefont{R{\o}nnow}},
  \bibinfo{author}{\bibfnamefont{B.}~\bibnamefont{Lake}},
  \bibinfo{author}{\bibfnamefont{S.~M.} \bibnamefont{Hayden}},
  \bibinfo{author}{\bibfnamefont{G.}~\bibnamefont{Aeppli}},
  \bibinfo{author}{\bibfnamefont{T.~G.} \bibnamefont{Perring}},
  \bibinfo{author}{\bibfnamefont{M.}~\bibnamefont{Mangkorntong}},
  \bibinfo{author}{\bibfnamefont{M.}~\bibnamefont{Nohara}}, \bibnamefont{and}
  \bibinfo{author}{\bibfnamefont{H.}~\bibnamefont{Tagaki}},
  \bibinfo{journal}{Phys. Rev. Lett.} \textbf{\bibinfo{volume}{93}},
  \bibinfo{pages}{147002} (\bibinfo{year}{2004}).

\bibitem[{\citenamefont{Vignolle et~al.}(2005)\citenamefont{Vignolle, Hayden,
  McMorrow, {R{\o}nnow}, Lake, Frost, and Perring}}]{vign07}
\bibinfo{author}{\bibfnamefont{B.}~\bibnamefont{Vignolle}},
  \bibinfo{author}{\bibfnamefont{S.~M.} \bibnamefont{Hayden}},
  \bibinfo{author}{\bibfnamefont{D.~F.} \bibnamefont{McMorrow}},
  \bibinfo{author}{\bibfnamefont{H.~M.} \bibnamefont{{R{\o}nnow}}},
  \bibinfo{author}{\bibfnamefont{B.}~\bibnamefont{Lake}},
  \bibinfo{author}{\bibfnamefont{C.~D.} \bibnamefont{Frost}}, \bibnamefont{and}
  \bibinfo{author}{\bibfnamefont{T.~G.} \bibnamefont{Perring}}
  (\bibinfo{year}{2005}), \eprint{cond-mat/0701151}.

\bibitem[{\citenamefont{Hayden et~al.}(2004)\citenamefont{Hayden, Mook, Dai,
  Perring, and Do\u{g}an}}]{hayd04}
\bibinfo{author}{\bibfnamefont{S.~M.} \bibnamefont{Hayden}},
  \bibinfo{author}{\bibfnamefont{H.~A.} \bibnamefont{Mook}},
  \bibinfo{author}{\bibfnamefont{P.}~\bibnamefont{Dai}},
  \bibinfo{author}{\bibfnamefont{T.~G.} \bibnamefont{Perring}},
  \bibnamefont{and}
  \bibinfo{author}{\bibfnamefont{F.}~\bibnamefont{Do\u{g}an}},
  \bibinfo{journal}{Nature} \textbf{\bibinfo{volume}{429}},
  \bibinfo{pages}{531} (\bibinfo{year}{2004}).

\bibitem[{\citenamefont{Reznik et~al.}(2004)\citenamefont{Reznik, Bourges,
  Pintschovius, Endoh, Sidis, Matsui, and Tajima}}]{rezn04}
\bibinfo{author}{\bibfnamefont{D.}~\bibnamefont{Reznik}},
  \bibinfo{author}{\bibfnamefont{P.}~\bibnamefont{Bourges}},
  \bibinfo{author}{\bibfnamefont{L.}~\bibnamefont{Pintschovius}},
  \bibinfo{author}{\bibfnamefont{Y.}~\bibnamefont{Endoh}},
  \bibinfo{author}{\bibfnamefont{Y.}~\bibnamefont{Sidis}},
  \bibinfo{author}{\bibfnamefont{T.}~\bibnamefont{Matsui}}, \bibnamefont{and}
  \bibinfo{author}{\bibfnamefont{S.}~\bibnamefont{Tajima}},
  \bibinfo{journal}{Phys. Rev. Lett.} \textbf{\bibinfo{volume}{93}},
  \bibinfo{pages}{207003} (\bibinfo{year}{2004}).

\bibitem[{\citenamefont{Stock et~al.}(2005)\citenamefont{Stock, Buyers, Cowley,
  Clegg, Coldea, Frost, Liang, Peets, Bonn, Hardy et~al.}}]{stoc05}
\bibinfo{author}{\bibfnamefont{C.}~\bibnamefont{Stock}},
  \bibinfo{author}{\bibfnamefont{W.~J.~L.} \bibnamefont{Buyers}},
  \bibinfo{author}{\bibfnamefont{R.~A.} \bibnamefont{Cowley}},
  \bibinfo{author}{\bibfnamefont{P.~S.} \bibnamefont{Clegg}},
  \bibinfo{author}{\bibfnamefont{R.}~\bibnamefont{Coldea}},
  \bibinfo{author}{\bibfnamefont{C.~D.} \bibnamefont{Frost}},
  \bibinfo{author}{\bibfnamefont{R.}~\bibnamefont{Liang}},
  \bibinfo{author}{\bibfnamefont{D.}~\bibnamefont{Peets}},
  \bibinfo{author}{\bibfnamefont{D.}~\bibnamefont{Bonn}},
  \bibinfo{author}{\bibfnamefont{W.~N.} \bibnamefont{Hardy}},
  \bibnamefont{et~al.}, \bibinfo{journal}{Phys. Rev. B}
  \textbf{\bibinfo{volume}{71}}, \bibinfo{pages}{024522}
  (\bibinfo{year}{2005}).

\bibitem[{\citenamefont{Pailh\`es et~al.}(2004)\citenamefont{Pailh\`es, Sidis,
  Bourges, Hinkov, Ivanov, Ulrich, Regnault, and Keimer}}]{pail04}
\bibinfo{author}{\bibfnamefont{S.}~\bibnamefont{Pailh\`es}},
  \bibinfo{author}{\bibfnamefont{Y.}~\bibnamefont{Sidis}},
  \bibinfo{author}{\bibfnamefont{P.}~\bibnamefont{Bourges}},
  \bibinfo{author}{\bibfnamefont{V.}~\bibnamefont{Hinkov}},
  \bibinfo{author}{\bibfnamefont{A.}~\bibnamefont{Ivanov}},
  \bibinfo{author}{\bibfnamefont{C.}~\bibnamefont{Ulrich}},
  \bibinfo{author}{\bibfnamefont{L.~P.} \bibnamefont{Regnault}},
  \bibnamefont{and} \bibinfo{author}{\bibfnamefont{B.}~\bibnamefont{Keimer}},
  \bibinfo{journal}{Phys. Rev. Lett.} \textbf{\bibinfo{volume}{93}},
  \bibinfo{pages}{167001} (\bibinfo{year}{2004}).

\bibitem[{\citenamefont{Hinkov et~al.}(2004)\citenamefont{Hinkov, Pailh\`{e}s,
  Bourges, Sidis, Ivanov, Kulakov, Lin, Chen, Bernhard, and Keimer}}]{hink04}
\bibinfo{author}{\bibfnamefont{V.}~\bibnamefont{Hinkov}},
  \bibinfo{author}{\bibfnamefont{S.}~\bibnamefont{Pailh\`{e}s}},
  \bibinfo{author}{\bibfnamefont{P.}~\bibnamefont{Bourges}},
  \bibinfo{author}{\bibfnamefont{Y.}~\bibnamefont{Sidis}},
  \bibinfo{author}{\bibfnamefont{A.}~\bibnamefont{Ivanov}},
  \bibinfo{author}{\bibfnamefont{A.}~\bibnamefont{Kulakov}},
  \bibinfo{author}{\bibfnamefont{C.~T.} \bibnamefont{Lin}},
  \bibinfo{author}{\bibfnamefont{D.~P.} \bibnamefont{Chen}},
  \bibinfo{author}{\bibfnamefont{C.}~\bibnamefont{Bernhard}}, \bibnamefont{and}
  \bibinfo{author}{\bibfnamefont{B.}~\bibnamefont{Keimer}},
  \bibinfo{journal}{Nature} \textbf{\bibinfo{volume}{430}},
  \bibinfo{pages}{650} (\bibinfo{year}{2004}).

\bibitem[{\citenamefont{Tranquada et~al.}(2006)\citenamefont{Tranquada, Woo,
  Perring, Goka, Gu, Xu, Fujita, and Yamada}}]{tran06a}
\bibinfo{author}{\bibfnamefont{J.~M.} \bibnamefont{Tranquada}},
  \bibinfo{author}{\bibfnamefont{H.}~\bibnamefont{Woo}},
  \bibinfo{author}{\bibfnamefont{T.~G.} \bibnamefont{Perring}},
  \bibinfo{author}{\bibfnamefont{H.}~\bibnamefont{Goka}},
  \bibinfo{author}{\bibfnamefont{G.~D.} \bibnamefont{Gu}},
  \bibinfo{author}{\bibfnamefont{G.}~\bibnamefont{Xu}},
  \bibinfo{author}{\bibfnamefont{M.}~\bibnamefont{Fujita}}, \bibnamefont{and}
  \bibinfo{author}{\bibfnamefont{K.}~\bibnamefont{Yamada}},
  \bibinfo{journal}{J. Phys. Chem. Solids} \textbf{\bibinfo{volume}{67}},
  \bibinfo{pages}{511} (\bibinfo{year}{2006}).

\bibitem[{\citenamefont{Bourges}(1998)}]{bour98}
\bibinfo{author}{\bibfnamefont{P.}~\bibnamefont{Bourges}}, in
  \emph{\bibinfo{booktitle}{The Gap Symmetry and Fluctuations in High
  Temperature Superconductors}}, edited by
  \bibinfo{editor}{\bibfnamefont{J.}~\bibnamefont{Bok}},
  \bibinfo{editor}{\bibfnamefont{G.}~\bibnamefont{Deutscher}},
  \bibinfo{editor}{\bibfnamefont{D.}~\bibnamefont{Pavuna}}, \bibnamefont{and}
  \bibinfo{editor}{\bibfnamefont{S.~A.} \bibnamefont{Wolf}}
  (\bibinfo{publisher}{Plenum}, \bibinfo{address}{New York},
  \bibinfo{year}{1998}), p. \bibinfo{pages}{349}.

\bibitem[{\citenamefont{Dai et~al.}(2001)\citenamefont{Dai, Mook, Hunt, and
  Do\u{g}an}}]{dai01}
\bibinfo{author}{\bibfnamefont{P.}~\bibnamefont{Dai}},
  \bibinfo{author}{\bibfnamefont{H.~A.} \bibnamefont{Mook}},
  \bibinfo{author}{\bibfnamefont{R.~D.} \bibnamefont{Hunt}}, \bibnamefont{and}
  \bibinfo{author}{\bibfnamefont{F.}~\bibnamefont{Do\u{g}an}},
  \bibinfo{journal}{Phys. Rev. B} \textbf{\bibinfo{volume}{63}},
  \bibinfo{pages}{054525} (\bibinfo{year}{2001}).

\bibitem[{\citenamefont{Fong et~al.}(1999)\citenamefont{Fong, Bourges, Sidis,
  Regnault, Ivanov, Gu, Koshizuka, and Keimer}}]{fong99}
\bibinfo{author}{\bibfnamefont{H.~F.} \bibnamefont{Fong}},
  \bibinfo{author}{\bibfnamefont{P.}~\bibnamefont{Bourges}},
  \bibinfo{author}{\bibfnamefont{Y.}~\bibnamefont{Sidis}},
  \bibinfo{author}{\bibfnamefont{L.~P.} \bibnamefont{Regnault}},
  \bibinfo{author}{\bibfnamefont{A.}~\bibnamefont{Ivanov}},
  \bibinfo{author}{\bibfnamefont{G.~D.} \bibnamefont{Gu}},
  \bibinfo{author}{\bibfnamefont{N.}~\bibnamefont{Koshizuka}},
  \bibnamefont{and} \bibinfo{author}{\bibfnamefont{B.}~\bibnamefont{Keimer}},
  \bibinfo{journal}{Nature} \textbf{\bibinfo{volume}{398}},
  \bibinfo{pages}{588} (\bibinfo{year}{1999}).

\bibitem[{\citenamefont{Tranquada}()}]{tran05c}
\bibinfo{author}{\bibfnamefont{J.~M.} \bibnamefont{Tranquada}},
  \bibinfo{note}{cond-mat/0512115}.

\bibitem[{\citenamefont{Tranquada
  et~al.}(2004{\natexlab{b}})\citenamefont{Tranquada, Woo, Perring, Goka, Gu,
  Xu, Fujita, and Yamada}}]{tran04}
\bibinfo{author}{\bibfnamefont{J.~M.} \bibnamefont{Tranquada}},
  \bibinfo{author}{\bibfnamefont{H.}~\bibnamefont{Woo}},
  \bibinfo{author}{\bibfnamefont{T.~G.} \bibnamefont{Perring}},
  \bibinfo{author}{\bibfnamefont{H.}~\bibnamefont{Goka}},
  \bibinfo{author}{\bibfnamefont{G.~D.} \bibnamefont{Gu}},
  \bibinfo{author}{\bibfnamefont{G.}~\bibnamefont{Xu}},
  \bibinfo{author}{\bibfnamefont{M.}~\bibnamefont{Fujita}}, \bibnamefont{and}
  \bibinfo{author}{\bibfnamefont{K.}~\bibnamefont{Yamada}},
  \bibinfo{journal}{Nature} \textbf{\bibinfo{volume}{429}},
  \bibinfo{pages}{534} (\bibinfo{year}{2004}{\natexlab{b}}).

\bibitem[{\citenamefont{Fujita et~al.}(2004)\citenamefont{Fujita, Goka, Yamada,
  Tranquada, and Regnault}}]{fuji04}
\bibinfo{author}{\bibfnamefont{M.}~\bibnamefont{Fujita}},
  \bibinfo{author}{\bibfnamefont{H.}~\bibnamefont{Goka}},
  \bibinfo{author}{\bibfnamefont{K.}~\bibnamefont{Yamada}},
  \bibinfo{author}{\bibfnamefont{J.~M.} \bibnamefont{Tranquada}},
  \bibnamefont{and} \bibinfo{author}{\bibfnamefont{L.~P.}
  \bibnamefont{Regnault}}, \bibinfo{journal}{Phys. Rev. B}
  \textbf{\bibinfo{volume}{70}}, \bibinfo{pages}{104517}
  (\bibinfo{year}{2004}).

\bibitem[{\citenamefont{Abbamonte et~al.}(2005)\citenamefont{Abbamonte, Rusydi,
  Smadici, Gu, Sawatzky, and Feng}}]{abba05}
\bibinfo{author}{\bibfnamefont{P.}~\bibnamefont{Abbamonte}},
  \bibinfo{author}{\bibfnamefont{A.}~\bibnamefont{Rusydi}},
  \bibinfo{author}{\bibfnamefont{S.}~\bibnamefont{Smadici}},
  \bibinfo{author}{\bibfnamefont{G.~D.} \bibnamefont{Gu}},
  \bibinfo{author}{\bibfnamefont{G.~A.} \bibnamefont{Sawatzky}},
  \bibnamefont{and} \bibinfo{author}{\bibfnamefont{D.~L.} \bibnamefont{Feng}},
  \bibinfo{journal}{Nature Phys.} \textbf{\bibinfo{volume}{1}},
  \bibinfo{pages}{155} (\bibinfo{year}{2005}).

\bibitem[{\citenamefont{H\"ucker and v.~Zimmermann}()}]{huckun}
\bibinfo{author}{\bibfnamefont{M.}~\bibnamefont{H\"ucker}} \bibnamefont{and}
  \bibinfo{author}{\bibfnamefont{M.}~\bibnamefont{v.~Zimmermann}},
  \bibinfo{note}{(unpublished)}.

\bibitem[{\citenamefont{H\"ucker et~al.}(2005)\citenamefont{H\"ucker, Gu, and
  Tranquada}}]{huck05b}
\bibinfo{author}{\bibfnamefont{M.}~\bibnamefont{H\"ucker}},
  \bibinfo{author}{\bibfnamefont{G.~D.} \bibnamefont{Gu}}, \bibnamefont{and}
  \bibinfo{author}{\bibfnamefont{J.~M.} \bibnamefont{Tranquada}}
  (\bibinfo{year}{2005}), \eprint{cond-mat/0503417}.

\bibitem[{\citenamefont{{Savici et al.}}(2005)}]{savi05}
\bibinfo{author}{\bibfnamefont{A.~T.} \bibnamefont{{Savici et al.}}},
  \bibinfo{journal}{Phys. Rev. Lett.} \textbf{\bibinfo{volume}{95}},
  \bibinfo{pages}{157001} (\bibinfo{year}{2005}).

\bibitem[{\citenamefont{Boothroyd et~al.}(2003)\citenamefont{Boothroyd,
  Prabhakaran, Freeman, Lister, Enderle, Hiess, and Kulda}}]{boot03}
\bibinfo{author}{\bibfnamefont{A.~T.} \bibnamefont{Boothroyd}},
  \bibinfo{author}{\bibfnamefont{D.}~\bibnamefont{Prabhakaran}},
  \bibinfo{author}{\bibfnamefont{P.~G.} \bibnamefont{Freeman}},
  \bibinfo{author}{\bibfnamefont{S.~J.~S.} \bibnamefont{Lister}},
  \bibinfo{author}{\bibfnamefont{M.}~\bibnamefont{Enderle}},
  \bibinfo{author}{\bibfnamefont{A.}~\bibnamefont{Hiess}}, \bibnamefont{and}
  \bibinfo{author}{\bibfnamefont{J.}~\bibnamefont{Kulda}},
  \bibinfo{journal}{Phys. Rev. B} \textbf{\bibinfo{volume}{67}},
  \bibinfo{pages}{100407(R)} (\bibinfo{year}{2003}).

\bibitem[{\citenamefont{Bourges et~al.}(2003)\citenamefont{Bourges, Sidis,
  Braden, Nakajima, and Tranquada}}]{bour03}
\bibinfo{author}{\bibfnamefont{P.}~\bibnamefont{Bourges}},
  \bibinfo{author}{\bibfnamefont{Y.}~\bibnamefont{Sidis}},
  \bibinfo{author}{\bibfnamefont{M.}~\bibnamefont{Braden}},
  \bibinfo{author}{\bibfnamefont{K.}~\bibnamefont{Nakajima}}, \bibnamefont{and}
  \bibinfo{author}{\bibfnamefont{J.~M.} \bibnamefont{Tranquada}},
  \bibinfo{journal}{Phys. Rev. Lett.} \textbf{\bibinfo{volume}{90}},
  \bibinfo{pages}{147202} (\bibinfo{year}{2003}).

\bibitem[{\citenamefont{Woo et~al.}(2005)\citenamefont{Woo, Boothroyd,
  Nakajima, Perring, Frost, Freeman, Prabhakaran, Yamada, and
  Tranquada}}]{woo05}
\bibinfo{author}{\bibfnamefont{H.}~\bibnamefont{Woo}},
  \bibinfo{author}{\bibfnamefont{A.~T.} \bibnamefont{Boothroyd}},
  \bibinfo{author}{\bibfnamefont{K.}~\bibnamefont{Nakajima}},
  \bibinfo{author}{\bibfnamefont{T.~G.} \bibnamefont{Perring}},
  \bibinfo{author}{\bibfnamefont{C.~D.} \bibnamefont{Frost}},
  \bibinfo{author}{\bibfnamefont{P.~G.} \bibnamefont{Freeman}},
  \bibinfo{author}{\bibfnamefont{D.}~\bibnamefont{Prabhakaran}},
  \bibinfo{author}{\bibfnamefont{K.}~\bibnamefont{Yamada}}, \bibnamefont{and}
  \bibinfo{author}{\bibfnamefont{J.~M.} \bibnamefont{Tranquada}},
  \bibinfo{journal}{Phys. Rev. B} \textbf{\bibinfo{volume}{72}},
  \bibinfo{pages}{064437} (\bibinfo{year}{2005}).

\bibitem[{\citenamefont{Homes et~al.}(2006)\citenamefont{Homes, Dordevic, Gu,
  Li, Valla, and Tranquada}}]{home06}
\bibinfo{author}{\bibfnamefont{C.~C.} \bibnamefont{Homes}},
  \bibinfo{author}{\bibfnamefont{S.~V.} \bibnamefont{Dordevic}},
  \bibinfo{author}{\bibfnamefont{G.~D.} \bibnamefont{Gu}},
  \bibinfo{author}{\bibfnamefont{Q.}~\bibnamefont{Li}},
  \bibinfo{author}{\bibfnamefont{T.}~\bibnamefont{Valla}}, \bibnamefont{and}
  \bibinfo{author}{\bibfnamefont{J.~M.} \bibnamefont{Tranquada}},
  \bibinfo{journal}{Phys. Rev. Lett.} \textbf{\bibinfo{volume}{96}},
  \bibinfo{pages}{257002} (\bibinfo{year}{2006}).

\bibitem[{\citenamefont{Valla et~al.}(2006)\citenamefont{Valla, Federov, Lee,
  Davis, and Gu}}]{vall06}
\bibinfo{author}{\bibfnamefont{T.}~\bibnamefont{Valla}},
  \bibinfo{author}{\bibfnamefont{A.~V.} \bibnamefont{Federov}},
  \bibinfo{author}{\bibfnamefont{J.}~\bibnamefont{Lee}},
  \bibinfo{author}{\bibfnamefont{J.~C.} \bibnamefont{Davis}}, \bibnamefont{and}
  \bibinfo{author}{\bibfnamefont{G.~D.} \bibnamefont{Gu}},
  \bibinfo{journal}{Science} \textbf{\bibinfo{volume}{314}},
  \bibinfo{pages}{1914} (\bibinfo{year}{2006}).

\bibitem[{\citenamefont{Shamoto et~al.}(1993)\citenamefont{Shamoto, Sato,
  Tranquada, Sternlieb, and Shirane}}]{sham93}
\bibinfo{author}{\bibfnamefont{S.}~\bibnamefont{Shamoto}},
  \bibinfo{author}{\bibfnamefont{M.}~\bibnamefont{Sato}},
  \bibinfo{author}{\bibfnamefont{J.~M.} \bibnamefont{Tranquada}},
  \bibinfo{author}{\bibfnamefont{B.~J.} \bibnamefont{Sternlieb}},
  \bibnamefont{and} \bibinfo{author}{\bibfnamefont{G.}~\bibnamefont{Shirane}},
  \bibinfo{journal}{Phys. Rev. B} \textbf{\bibinfo{volume}{48}},
  \bibinfo{pages}{13817} (\bibinfo{year}{1993}).

\bibitem[{\citenamefont{Tranquada et~al.}(1995)\citenamefont{Tranquada,
  Sternlieb, Axe, Nakamura, and Uchida}}]{tran95a}
\bibinfo{author}{\bibfnamefont{J.~M.} \bibnamefont{Tranquada}},
  \bibinfo{author}{\bibfnamefont{B.~J.} \bibnamefont{Sternlieb}},
  \bibinfo{author}{\bibfnamefont{J.~D.} \bibnamefont{Axe}},
  \bibinfo{author}{\bibfnamefont{Y.}~\bibnamefont{Nakamura}}, \bibnamefont{and}
  \bibinfo{author}{\bibfnamefont{S.}~\bibnamefont{Uchida}},
  \bibinfo{journal}{Nature} \textbf{\bibinfo{volume}{375}},
  \bibinfo{pages}{561} (\bibinfo{year}{1995}).

\bibitem[{\citenamefont{Yamada et~al.}(1989)\citenamefont{Yamada, Kakurai,
  Endoh, Thurston, Kastner, Birgeneau, Shirane, Hidaka, and Murakami}}]{yama89}
\bibinfo{author}{\bibfnamefont{K.}~\bibnamefont{Yamada}},
  \bibinfo{author}{\bibfnamefont{K.}~\bibnamefont{Kakurai}},
  \bibinfo{author}{\bibfnamefont{Y.}~\bibnamefont{Endoh}},
  \bibinfo{author}{\bibfnamefont{T.~R.} \bibnamefont{Thurston}},
  \bibinfo{author}{\bibfnamefont{M.~A.} \bibnamefont{Kastner}},
  \bibinfo{author}{\bibfnamefont{R.~J.} \bibnamefont{Birgeneau}},
  \bibinfo{author}{\bibfnamefont{G.}~\bibnamefont{Shirane}},
  \bibinfo{author}{\bibfnamefont{Y.}~\bibnamefont{Hidaka}}, \bibnamefont{and}
  \bibinfo{author}{\bibfnamefont{T.}~\bibnamefont{Murakami}},
  \bibinfo{journal}{Phys. Rev. B} \textbf{\bibinfo{volume}{40}},
  \bibinfo{pages}{4557} (\bibinfo{year}{1989}).

\bibitem[{\citenamefont{Aeppli et~al.}(1997)\citenamefont{Aeppli, Mason,
  Hayden, Mook, and Kulda}}]{aepp97}
\bibinfo{author}{\bibfnamefont{G.}~\bibnamefont{Aeppli}},
  \bibinfo{author}{\bibfnamefont{T.~E.} \bibnamefont{Mason}},
  \bibinfo{author}{\bibfnamefont{S.~M.} \bibnamefont{Hayden}},
  \bibinfo{author}{\bibfnamefont{H.~A.} \bibnamefont{Mook}}, \bibnamefont{and}
  \bibinfo{author}{\bibfnamefont{J.}~\bibnamefont{Kulda}},
  \bibinfo{journal}{Science} \textbf{\bibinfo{volume}{278}},
  \bibinfo{pages}{1432} (\bibinfo{year}{1997}).

\bibitem[{\citenamefont{Wakimoto et~al.}(2006)\citenamefont{Wakimoto, Yamada,
  Tranquada, Frost, Birgeneau, and Zhang}}]{waki06}
\bibinfo{author}{\bibfnamefont{S.}~\bibnamefont{Wakimoto}},
  \bibinfo{author}{\bibfnamefont{K.}~\bibnamefont{Yamada}},
  \bibinfo{author}{\bibfnamefont{J.~M.} \bibnamefont{Tranquada}},
  \bibinfo{author}{\bibfnamefont{C.~D.} \bibnamefont{Frost}},
  \bibinfo{author}{\bibfnamefont{R.~J.} \bibnamefont{Birgeneau}},
  \bibnamefont{and} \bibinfo{author}{\bibfnamefont{H.}~\bibnamefont{Zhang}}
  (\bibinfo{year}{2006}), \eprint{cond-mat/0609155}.

\bibitem[{\citenamefont{Hayden et~al.}(1996)\citenamefont{Hayden, Aeppli, Mook,
  Perring, Mason, Cheong, and Fisk}}]{hayd96a}
\bibinfo{author}{\bibfnamefont{S.~M.} \bibnamefont{Hayden}},
  \bibinfo{author}{\bibfnamefont{G.}~\bibnamefont{Aeppli}},
  \bibinfo{author}{\bibfnamefont{H.~A.} \bibnamefont{Mook}},
  \bibinfo{author}{\bibfnamefont{T.~G.} \bibnamefont{Perring}},
  \bibinfo{author}{\bibfnamefont{T.~E.} \bibnamefont{Mason}},
  \bibinfo{author}{\bibfnamefont{S.-W.} \bibnamefont{Cheong}},
  \bibnamefont{and} \bibinfo{author}{\bibfnamefont{Z.}~\bibnamefont{Fisk}},
  \bibinfo{journal}{Phys. Rev. Lett.} \textbf{\bibinfo{volume}{76}},
  \bibinfo{pages}{1344} (\bibinfo{year}{1996}).

\bibitem[{\citenamefont{Zaanen et~al.}(1996)\citenamefont{Zaanen, Horbach, and
  van Saarloos}}]{zaan96a}
\bibinfo{author}{\bibfnamefont{J.}~\bibnamefont{Zaanen}},
  \bibinfo{author}{\bibfnamefont{M.~L.} \bibnamefont{Horbach}},
  \bibnamefont{and} \bibinfo{author}{\bibfnamefont{W.}~\bibnamefont{van
  Saarloos}}, \bibinfo{journal}{Phys. Rev. B} \textbf{\bibinfo{volume}{53}},
  \bibinfo{pages}{8671} (\bibinfo{year}{1996}).

\bibitem[{\citenamefont{Kivelson et~al.}(1998)\citenamefont{Kivelson, Fradkin,
  and Emery}}]{kive98}
\bibinfo{author}{\bibfnamefont{S.~A.} \bibnamefont{Kivelson}},
  \bibinfo{author}{\bibfnamefont{E.}~\bibnamefont{Fradkin}}, \bibnamefont{and}
  \bibinfo{author}{\bibfnamefont{V.~J.} \bibnamefont{Emery}},
  \bibinfo{journal}{Nature} \textbf{\bibinfo{volume}{393}},
  \bibinfo{pages}{550} (\bibinfo{year}{1998}).

\bibitem[{\citenamefont{Lee et~al.}(2002)\citenamefont{Lee, Tranquada, Yamada,
  Buttrey, Li, and Cheong}}]{lee02}
\bibinfo{author}{\bibfnamefont{S.-H.} \bibnamefont{Lee}},
  \bibinfo{author}{\bibfnamefont{J.~M.} \bibnamefont{Tranquada}},
  \bibinfo{author}{\bibfnamefont{K.}~\bibnamefont{Yamada}},
  \bibinfo{author}{\bibfnamefont{D.~J.} \bibnamefont{Buttrey}},
  \bibinfo{author}{\bibfnamefont{Q.}~\bibnamefont{Li}}, \bibnamefont{and}
  \bibinfo{author}{\bibfnamefont{S.-W.} \bibnamefont{Cheong}},
  \bibinfo{journal}{Phys. Rev. Lett.} \textbf{\bibinfo{volume}{88}},
  \bibinfo{pages}{126401} (\bibinfo{year}{2002}).

\bibitem[{\citenamefont{Vojta and Ulbricht}(2004)}]{vojt04}
\bibinfo{author}{\bibfnamefont{M.}~\bibnamefont{Vojta}} \bibnamefont{and}
  \bibinfo{author}{\bibfnamefont{T.}~\bibnamefont{Ulbricht}},
  \bibinfo{journal}{Phys. Rev. Lett.} \textbf{\bibinfo{volume}{93}},
  \bibinfo{pages}{127002} (\bibinfo{year}{2004}).

\bibitem[{\citenamefont{Uhrig et~al.}(2004)\citenamefont{Uhrig, Schmidt, and
  Gr\"uninger}}]{uhri04}
\bibinfo{author}{\bibfnamefont{G.~S.} \bibnamefont{Uhrig}},
  \bibinfo{author}{\bibfnamefont{K.~P.} \bibnamefont{Schmidt}},
  \bibnamefont{and}
  \bibinfo{author}{\bibfnamefont{M.}~\bibnamefont{Gr\"uninger}},
  \bibinfo{journal}{Phys. Rev. Lett.} \textbf{\bibinfo{volume}{93}},
  \bibinfo{pages}{267003} (\bibinfo{year}{2004}).

\bibitem[{\citenamefont{Vojta et~al.}(2006)\citenamefont{Vojta, Vojta, and
  Kaul}}]{vojt06}
\bibinfo{author}{\bibfnamefont{M.}~\bibnamefont{Vojta}},
  \bibinfo{author}{\bibfnamefont{T.}~\bibnamefont{Vojta}}, \bibnamefont{and}
  \bibinfo{author}{\bibfnamefont{R.~K.} \bibnamefont{Kaul}},
  \bibinfo{journal}{Phys. Rev. Lett.} \textbf{\bibinfo{volume}{97}},
  \bibinfo{pages}{097001} (\bibinfo{year}{2006}).

\bibitem[{\citenamefont{Yao et~al.}(2006)\citenamefont{Yao, Carlson, and
  Campbell}}]{yao06}
\bibinfo{author}{\bibfnamefont{D.~X.} \bibnamefont{Yao}},
  \bibinfo{author}{\bibfnamefont{E.~W.} \bibnamefont{Carlson}},
  \bibnamefont{and} \bibinfo{author}{\bibfnamefont{D.~K.}
  \bibnamefont{Campbell}}, \bibinfo{journal}{Phys. Rev. Lett.}
  \textbf{\bibinfo{volume}{97}}, \bibinfo{pages}{017003}
  (\bibinfo{year}{2006}).

\bibitem[{\citenamefont{Andersen and Sylju{\aa}sen}(2006)}]{ande06}
\bibinfo{author}{\bibfnamefont{B.~M.} \bibnamefont{Andersen}} \bibnamefont{and}
  \bibinfo{author}{\bibfnamefont{O.~F.} \bibnamefont{Sylju{\aa}sen}}
  (\bibinfo{year}{2006}), \eprint{cond-mat/0608251}.

\bibitem[{\citenamefont{Hinkov et~al.}(2006)\citenamefont{Hinkov, Bourges,
  Pailh{\`e}s, Sidis, Ivanov, Lin, Chen, and Keimer}}]{hink06}
\bibinfo{author}{\bibfnamefont{V.}~\bibnamefont{Hinkov}},
  \bibinfo{author}{\bibfnamefont{P.}~\bibnamefont{Bourges}},
  \bibinfo{author}{\bibfnamefont{S.}~\bibnamefont{Pailh{\`e}s}},
  \bibinfo{author}{\bibfnamefont{Y.}~\bibnamefont{Sidis}},
  \bibinfo{author}{\bibfnamefont{A.}~\bibnamefont{Ivanov}},
  \bibinfo{author}{\bibfnamefont{C.~T.} \bibnamefont{Lin}},
  \bibinfo{author}{\bibfnamefont{D.~P.} \bibnamefont{Chen}}, \bibnamefont{and}
  \bibinfo{author}{\bibfnamefont{B.}~\bibnamefont{Keimer}}
  (\bibinfo{year}{2006}), \eprint{cond-mat/0601048}.

\bibitem[{\citenamefont{Piskunov et~al.}(2005)\citenamefont{Piskunov,
  {J\'erome}, Auban-Senzier, Wzietek, and Yakubovsky}}]{pisk05}
\bibinfo{author}{\bibfnamefont{Y.}~\bibnamefont{Piskunov}},
  \bibinfo{author}{\bibfnamefont{D.}~\bibnamefont{{J\'erome}}},
  \bibinfo{author}{\bibfnamefont{P.}~\bibnamefont{Auban-Senzier}},
  \bibinfo{author}{\bibfnamefont{P.}~\bibnamefont{Wzietek}}, \bibnamefont{and}
  \bibinfo{author}{\bibfnamefont{A.}~\bibnamefont{Yakubovsky}},
  \bibinfo{journal}{Phys. Rev. B} \textbf{\bibinfo{volume}{72}},
  \bibinfo{pages}{064512} (\bibinfo{year}{2005}).

\bibitem[{\citenamefont{{Vuleti\'c} et~al.}(2005)\citenamefont{{Vuleti\'c},
  Ivek, {Korin-Hamzi\'c}, {Tomi\'c}, Gorshunov, Dressel, Hess, {B\"uchner}, and
  Akimitsu}}]{vule05}
\bibinfo{author}{\bibfnamefont{T.}~\bibnamefont{{Vuleti\'c}}},
  \bibinfo{author}{\bibfnamefont{T.}~\bibnamefont{Ivek}},
  \bibinfo{author}{\bibfnamefont{B.}~\bibnamefont{{Korin-Hamzi\'c}}},
  \bibinfo{author}{\bibfnamefont{S.}~\bibnamefont{{Tomi\'c}}},
  \bibinfo{author}{\bibfnamefont{B.}~\bibnamefont{Gorshunov}},
  \bibinfo{author}{\bibfnamefont{M.}~\bibnamefont{Dressel}},
  \bibinfo{author}{\bibfnamefont{C.}~\bibnamefont{Hess}},
  \bibinfo{author}{\bibfnamefont{B.}~\bibnamefont{{B\"uchner}}},
  \bibnamefont{and} \bibinfo{author}{\bibfnamefont{J.}~\bibnamefont{Akimitsu}},
  \bibinfo{journal}{J. Phys. IV France} \textbf{\bibinfo{volume}{131}},
  \bibinfo{pages}{299} (\bibinfo{year}{2005}).

\bibitem[{\citenamefont{Rusydi et~al.}(2006)\citenamefont{Rusydi, Abbamonte,
  Berciu, Smadici, Eisaki, Fujimaki, Uchida, {R\"ubhausen}, and
  Sawatzky}}]{rusy06}
\bibinfo{author}{\bibfnamefont{A.}~\bibnamefont{Rusydi}},
  \bibinfo{author}{\bibfnamefont{P.}~\bibnamefont{Abbamonte}},
  \bibinfo{author}{\bibfnamefont{M.}~\bibnamefont{Berciu}},
  \bibinfo{author}{\bibfnamefont{S.}~\bibnamefont{Smadici}},
  \bibinfo{author}{\bibfnamefont{H.}~\bibnamefont{Eisaki}},
  \bibinfo{author}{\bibfnamefont{Y.}~\bibnamefont{Fujimaki}},
  \bibinfo{author}{\bibfnamefont{S.}~\bibnamefont{Uchida}},
  \bibinfo{author}{\bibfnamefont{M.}~\bibnamefont{{R\"ubhausen}}},
  \bibnamefont{and} \bibinfo{author}{\bibfnamefont{G.~A.}
  \bibnamefont{Sawatzky}} (\bibinfo{year}{2006}), \eprint{cond-mat/0604101}.

\bibitem[{\citenamefont{Boullier et~al.}(2004)\citenamefont{Boullier, Regnault,
  Lorenzo, R{\o}nnow, Ammerahl, Dhalenne, and Revcolevschi}}]{boul04}
\bibinfo{author}{\bibfnamefont{C.}~\bibnamefont{Boullier}},
  \bibinfo{author}{\bibfnamefont{L.~P.} \bibnamefont{Regnault}},
  \bibinfo{author}{\bibfnamefont{J.~E.} \bibnamefont{Lorenzo}},
  \bibinfo{author}{\bibfnamefont{H.~M.} \bibnamefont{R{\o}nnow}},
  \bibinfo{author}{\bibfnamefont{U.}~\bibnamefont{Ammerahl}},
  \bibinfo{author}{\bibfnamefont{G.}~\bibnamefont{Dhalenne}}, \bibnamefont{and}
  \bibinfo{author}{\bibfnamefont{A.}~\bibnamefont{Revcolevschi}},
  \bibinfo{journal}{Physica B} \textbf{\bibinfo{volume}{350}},
  \bibinfo{pages}{40} (\bibinfo{year}{2004}).

\bibitem[{\citenamefont{Notbohm et~al.}(2007)\citenamefont{Notbohm, Ribeiro,
  Lake, Tennant, Schmidt, Uhrig, Hess, Klingeler, Behr, {B\"uchner}
  et~al.}}]{notb07}
\bibinfo{author}{\bibfnamefont{S.}~\bibnamefont{Notbohm}},
  \bibinfo{author}{\bibfnamefont{P.}~\bibnamefont{Ribeiro}},
  \bibinfo{author}{\bibfnamefont{B.}~\bibnamefont{Lake}},
  \bibinfo{author}{\bibfnamefont{D.~A.} \bibnamefont{Tennant}},
  \bibinfo{author}{\bibfnamefont{K.~P.} \bibnamefont{Schmidt}},
  \bibinfo{author}{\bibfnamefont{G.~S.} \bibnamefont{Uhrig}},
  \bibinfo{author}{\bibfnamefont{C.}~\bibnamefont{Hess}},
  \bibinfo{author}{\bibfnamefont{R.}~\bibnamefont{Klingeler}},
  \bibinfo{author}{\bibfnamefont{G.}~\bibnamefont{Behr}},
  \bibinfo{author}{\bibfnamefont{B.}~\bibnamefont{{B\"uchner}}},
  \bibnamefont{et~al.}, \bibinfo{journal}{Phys. Rev. Lett.}
  \textbf{\bibinfo{volume}{98}}, \bibinfo{pages}{027403}
  (\bibinfo{year}{2007}).

\bibitem[{\citenamefont{Roux et~al.}(2005)\citenamefont{Roux, White, Poilblanc,
  Capponi, and {L\"auchli}}}]{roux05}
\bibinfo{author}{\bibfnamefont{G.}~\bibnamefont{Roux}},
  \bibinfo{author}{\bibfnamefont{S.~R.} \bibnamefont{White}},
  \bibinfo{author}{\bibfnamefont{D.}~\bibnamefont{Poilblanc}},
  \bibinfo{author}{\bibfnamefont{S.}~\bibnamefont{Capponi}}, \bibnamefont{and}
  \bibinfo{author}{\bibfnamefont{A.}~\bibnamefont{{L\"auchli}}},
  \bibinfo{journal}{Phys. Rev. B} \textbf{\bibinfo{volume}{72}},
  \bibinfo{pages}{014523} (\bibinfo{year}{2005}).

\bibitem[{\citenamefont{Essler and Konik}(2006)}]{essl06}
\bibinfo{author}{\bibfnamefont{F.~H.~L.} \bibnamefont{Essler}}
  \bibnamefont{and} \bibinfo{author}{\bibfnamefont{R.~M.} \bibnamefont{Konik}}
  (\bibinfo{year}{2006}), \eprint{cond-mat/0607783}.

\bibitem[{\citenamefont{Dagotto et~al.}(1992)\citenamefont{Dagotto, Riera, and
  Scalapino}}]{dago92}
\bibinfo{author}{\bibfnamefont{E.}~\bibnamefont{Dagotto}},
  \bibinfo{author}{\bibfnamefont{J.}~\bibnamefont{Riera}}, \bibnamefont{and}
  \bibinfo{author}{\bibfnamefont{D.~J.} \bibnamefont{Scalapino}},
  \bibinfo{journal}{Phys. Rev. B} \textbf{\bibinfo{volume}{45}},
  \bibinfo{pages}{5744} (\bibinfo{year}{1992}).

\bibitem[{\citenamefont{Kivelson and Fradkin}()}]{kive05a}
\bibinfo{author}{\bibfnamefont{S.~A.} \bibnamefont{Kivelson}} \bibnamefont{and}
  \bibinfo{author}{\bibfnamefont{E.}~\bibnamefont{Fradkin}},
  \bibinfo{note}{cond-mat/0507459}.

\bibitem[{\citenamefont{Xu et~al.}(1996)\citenamefont{Xu, DiTusa, Ito, Oka,
  Takagi, Broholm, and Aeppli}}]{xu96}
\bibinfo{author}{\bibfnamefont{G.}~\bibnamefont{Xu}},
  \bibinfo{author}{\bibfnamefont{J.~F.} \bibnamefont{DiTusa}},
  \bibinfo{author}{\bibfnamefont{T.}~\bibnamefont{Ito}},
  \bibinfo{author}{\bibfnamefont{K.}~\bibnamefont{Oka}},
  \bibinfo{author}{\bibfnamefont{H.}~\bibnamefont{Takagi}},
  \bibinfo{author}{\bibfnamefont{C.}~\bibnamefont{Broholm}}, \bibnamefont{and}
  \bibinfo{author}{\bibfnamefont{G.}~\bibnamefont{Aeppli}},
  \bibinfo{journal}{Phys. Rev. B} \textbf{\bibinfo{volume}{54}},
  \bibinfo{pages}{R6827} (\bibinfo{year}{1996}).

\bibitem[{\citenamefont{Xu et~al.}(2000)\citenamefont{Xu, Aeppli, Bisher,
  Broholm, DiTusa, Frost, Ito, Oka, Paul, Takagi et~al.}}]{xu00b}
\bibinfo{author}{\bibfnamefont{G.}~\bibnamefont{Xu}},
  \bibinfo{author}{\bibfnamefont{G.}~\bibnamefont{Aeppli}},
  \bibinfo{author}{\bibfnamefont{M.~E.} \bibnamefont{Bisher}},
  \bibinfo{author}{\bibfnamefont{C.}~\bibnamefont{Broholm}},
  \bibinfo{author}{\bibfnamefont{J.~F.} \bibnamefont{DiTusa}},
  \bibinfo{author}{\bibfnamefont{C.~D.} \bibnamefont{Frost}},
  \bibinfo{author}{\bibfnamefont{T.}~\bibnamefont{Ito}},
  \bibinfo{author}{\bibfnamefont{K.}~\bibnamefont{Oka}},
  \bibinfo{author}{\bibfnamefont{R.~L.} \bibnamefont{Paul}},
  \bibinfo{author}{\bibfnamefont{H.}~\bibnamefont{Takagi}},
  \bibnamefont{et~al.}, \bibinfo{journal}{Science}
  \textbf{\bibinfo{volume}{289}}, \bibinfo{pages}{419} (\bibinfo{year}{2000}).

\bibitem[{\citenamefont{Seibold and Lorenzana}(2005)}]{seib05}
\bibinfo{author}{\bibfnamefont{G.}~\bibnamefont{Seibold}} \bibnamefont{and}
  \bibinfo{author}{\bibfnamefont{J.}~\bibnamefont{Lorenzana}},
  \bibinfo{journal}{Phys. Rev. Lett.} \textbf{\bibinfo{volume}{94}},
  \bibinfo{pages}{107006} (\bibinfo{year}{2005}).

\bibitem[{\citenamefont{Bourges et~al.}(2000)\citenamefont{Bourges, Sidis,
  Fong, Regnault, Bossy, Ivanov, and Keimer}}]{bour00}
\bibinfo{author}{\bibfnamefont{P.}~\bibnamefont{Bourges}},
  \bibinfo{author}{\bibfnamefont{Y.}~\bibnamefont{Sidis}},
  \bibinfo{author}{\bibfnamefont{H.~F.} \bibnamefont{Fong}},
  \bibinfo{author}{\bibfnamefont{L.~P.} \bibnamefont{Regnault}},
  \bibinfo{author}{\bibfnamefont{J.}~\bibnamefont{Bossy}},
  \bibinfo{author}{\bibfnamefont{A.}~\bibnamefont{Ivanov}}, \bibnamefont{and}
  \bibinfo{author}{\bibfnamefont{B.}~\bibnamefont{Keimer}},
  \bibinfo{journal}{Science} \textbf{\bibinfo{volume}{288}},
  \bibinfo{pages}{1234} (\bibinfo{year}{2000}).

\bibitem[{\citenamefont{Gao et~al.}(1993)\citenamefont{Gao, Romero, Tanner,
  Talvacchio, and Forrester}}]{gao93}
\bibinfo{author}{\bibfnamefont{F.}~\bibnamefont{Gao}},
  \bibinfo{author}{\bibfnamefont{D.~B.} \bibnamefont{Romero}},
  \bibinfo{author}{\bibfnamefont{D.~B.} \bibnamefont{Tanner}},
  \bibinfo{author}{\bibfnamefont{J.}~\bibnamefont{Talvacchio}},
  \bibnamefont{and} \bibinfo{author}{\bibfnamefont{M.~G.}
  \bibnamefont{Forrester}}, \bibinfo{journal}{Phys. Rev. B}
  \textbf{\bibinfo{volume}{47}}, \bibinfo{pages}{1036} (\bibinfo{year}{1993}).

\bibitem[{\citenamefont{Takeya et~al.}(2002)\citenamefont{Takeya, Ando, Komiya,
  and Sun}}]{take02}
\bibinfo{author}{\bibfnamefont{J.}~\bibnamefont{Takeya}},
  \bibinfo{author}{\bibfnamefont{Y.}~\bibnamefont{Ando}},
  \bibinfo{author}{\bibfnamefont{S.}~\bibnamefont{Komiya}}, \bibnamefont{and}
  \bibinfo{author}{\bibfnamefont{X.~F.} \bibnamefont{Sun}},
  \bibinfo{journal}{Phys. Rev. Lett.} \textbf{\bibinfo{volume}{88}},
  \bibinfo{pages}{077001} (\bibinfo{year}{2002}).

\bibitem[{\citenamefont{Lee et~al.}(2005)\citenamefont{Lee, Segawa, Li,
  Padilla, Dumm, Dordevic, Homes, Ando, and Basov}}]{lee05}
\bibinfo{author}{\bibfnamefont{Y.~S.} \bibnamefont{Lee}},
  \bibinfo{author}{\bibfnamefont{K.}~\bibnamefont{Segawa}},
  \bibinfo{author}{\bibfnamefont{Z.~Q.} \bibnamefont{Li}},
  \bibinfo{author}{\bibfnamefont{W.~J.} \bibnamefont{Padilla}},
  \bibinfo{author}{\bibfnamefont{M.}~\bibnamefont{Dumm}},
  \bibinfo{author}{\bibfnamefont{S.~V.} \bibnamefont{Dordevic}},
  \bibinfo{author}{\bibfnamefont{C.~C.} \bibnamefont{Homes}},
  \bibinfo{author}{\bibfnamefont{Y.}~\bibnamefont{Ando}}, \bibnamefont{and}
  \bibinfo{author}{\bibfnamefont{D.~N.} \bibnamefont{Basov}},
  \bibinfo{journal}{Phys. Rev. B} \textbf{\bibinfo{volume}{72}},
  \bibinfo{pages}{054529} (\bibinfo{year}{2005}).

\end{thebibliography}

\end{document}